\documentclass[]{spie}  

\usepackage{amsmath,amsfonts,amssymb}
\usepackage{graphicx,verbatim}
\usepackage[colorlinks=true, allcolors=blue]{hyperref}

\title{Analysis of antenna position measurements and weather station network data during the ALMA Long Baseline Campaign of 2015}

\author[a]{Todd R. Hunter}
\author[b]{Robert Lucas}
\author[c]{Dominique Brogui{\`e}re}
\author[a,d]{Ed B. Fomalont}
\author[d]{William R. F. Dent}
\author[d]{Neil Phillips}
\author[d,e]{David Rabanus}
\author[a,d]{Catherine Vlahakis}
\affil[a]{National Radio Astronomy Observatory (NRAO), 520 Edgemont Rd, Charlottesville, VA, USA}
\affil[b]{Institut de Plan\'etologie et d'Astrophysique de Grenoble, Grenoble, France}
\affil[c]{Institut de Radioastronomie Millim\'etrique (IRAM), 300 rue de la Piscine, Domaine Universitaire, F-38406 Saint Martin d'H\'eres, France}
\affil[d]{Joint ALMA Observatory (JAO), Alonso de C\'ordova 3107, Vitacura, Santiago, Chile}
\affil[e]{APEX/European Southern Observatory, Alonso de C\'ordova 3107, Vitacura, Santiago, Chile}

\authorinfo{Further author information: (Send correspondence to T.R.H.)\\T.R.H: E-mail: thunter@nrao.edu, Telephone: 1 434 244 6836}

\begin{document} 
\maketitle

\begin{abstract}
In a radio interferometer, the geometrical antenna positions are determined from measurements of the observed delay to each antenna from observations across the sky of many point sources whose positions are known to high accuracy. The determination of accurate antenna positions relies on accurate calibration of the dry and wet delay of the atmosphere above each antenna. For the Atacama Large Millimeter/Submillimeter Array (ALMA), with baseline lengths up to 16 kilometers, the geography of the site forces the height above mean sea level of the more distant antenna pads to be significantly lower than the central array. Thus, both the ground level meteorological values and the total water column can be quite different between antennas in the extended configurations. During 2015, a network of six additional weather stations was installed to monitor pressure, temperature, relative humidity and wind velocity, in order to test whether inclusion of these parameters could improve the repeatability of antenna position determinations in these configurations.  We present an analysis of the data obtained during the ALMA Long Baseline Campaign of October through November 2015.  The repeatability of antenna position measurements typically degrades as a function of antenna distance.  Also, the scatter is more than three times worse in the vertical direction than in the local tangent plane, suggesting that a systematic effect is limiting the measurements.  So far we have explored correcting the delay model for deviations from hydrostatic equilibrium in the measured air pressure and separating the partial pressure of water from the total pressure using water vapor radiometer (WVR) data. Correcting for these combined effects still does not provide a good match to the residual position errors in the vertical direction.  One hypothesis is that the current model of water vapor may be too simple to fully remove the day-to-day variations in the wet delay.  We describe possible new avenues of improvement, which include recalibrating the baseline measurement datasets using the contemporaneous measurements of the water vapor scale height and temperature lapse rate from the oxygen sounder, and applying more accurate measurements of the sky coupling of the WVRs.
\end{abstract}

\keywords{Radio interferometry, water vapor radiometry, ALMA, antenna position determination}

\section{INTRODUCTION}
\label{sec:intro}  
In a radio interferometer, the geometrical antenna positions are determined from measurements of the observed delay to each antenna from observations across the sky of many point-source quasars whose positions have been established to milliarcsecond accuracy in the International Celestial Reference System (ICRS) \cite{Ma98}. The determination of geometrical antenna positions relies on accurate calibration of the dry and wet delay of the atmosphere above each antenna.  For ALMA \cite{Hills2010,Hills08}, the height of the pads above mean sea level varies significantly on the outer stations, which are arranged on three geographical branches: west, south and northeast.  Thus, the pressure measured at the central weather station must be scaled appropriately to each pad by the delay server component of the online control software \cite{Marson08,Amestica06,Farris05} while the WVRs\cite{Cherednichenko2010,Emrich09} provide a $\approx$1~Hz measurement of the time-variable wet path due to precipitable water vapor (PWV) which can be applied to the raw data as temporal phase corrections \cite{Nikolic13}.  During the initial ALMA Long Baseline Campaign (LBC) in 2014 \cite{LBC2014}, the solutions derived for the antenna positions on the distant stations by the online software module TelCal \cite{Telcal2011,Telcal2004} were found to vary significantly with time and/or weather conditions. In the end, the best position results were achieved by using the full wet path option of TelCal's {\tt tc\_antpos} task applied to the raw datastream uncorrected by the WVR measurements.  In other words, the visibility data that does not have the online WVR correction applied is used along with the full wet path inferred from the WVR data to determine the antenna positions most consistent with the measured delays across the sky.  Indeed, when the antenna positions determined in this manner were applied to the Science Verification data of the gravitational lens object \cite{SDP81}, it eliminated the need for a single phase-only self-calibration correction per execution to reach the best image quality.  Therefore, the plan going into the 2015 LBC (October 1 - November 30) was to use the "uncorrected, full wet path" option for determining antenna positions from the outset, while at the same time collecting additional weather data from the expanded network of six new weather stations deployed across the array to later determine whether this information could improve the consistency of the position measurements.  This report summarizes the work that was done during October-November 2015, and what research remains to be done.

\section{First series of antenna position measurements and corrections}

\begin{figure}[h!]   
\centering
\includegraphics[scale=0.62]{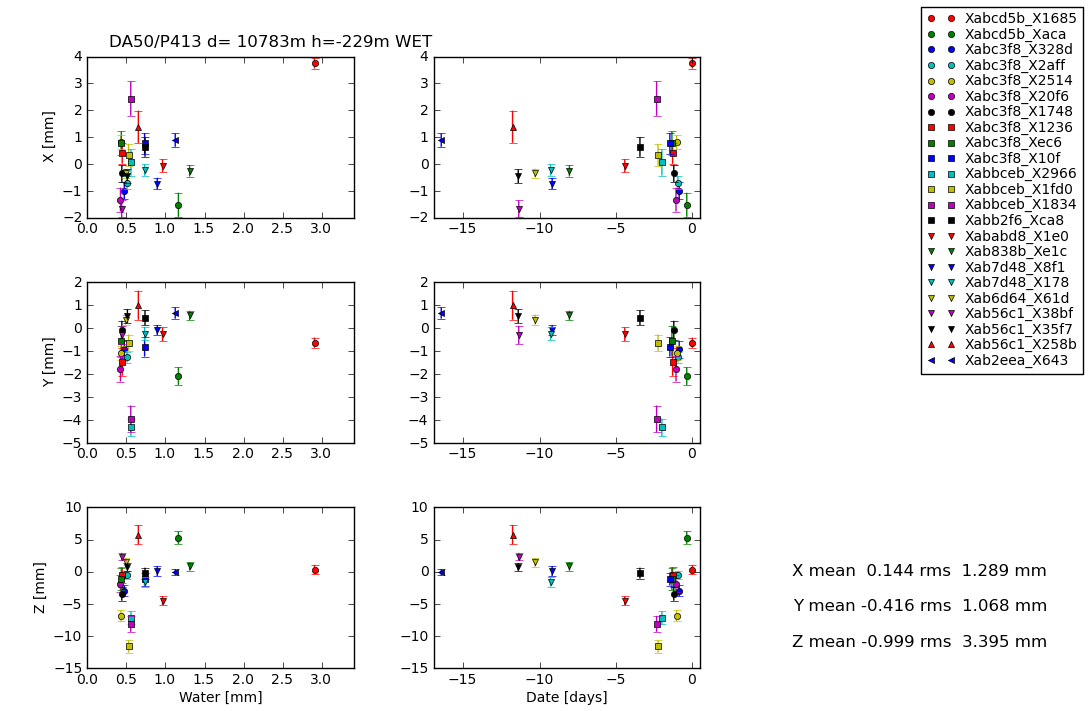}
\caption{Measured position corrections of DA50, located near the end of the Pampa La Bola branch, as a function of the PWV (left column) and observing date, where zero = 2015-Oct-16 (right column).  The mean and rms values listed in the lower right corner are weighted by the uncertainties in the individual measurements. }
\label{firstupdate}
\end{figure}

Antenna position measurements in the form of 40-minute observations of 50 quasars spread across the sky (``all-sky'' runs) were performed on most nights of the campaign. The first series of data was acquired between September 30 and October 16.  An example plot of the position obtained for a distant antenna (DA50) on the northeast branch, also called the Pampa La Bola branch (P branch) is shown in Figure~\ref{firstupdate}, which shows significant scatter in all three coordinates, but largest in the Z direction (local vertical). The position corrections for the more central antennas are generally more consistent than this example.  For each antenna, we identified a particular dataset whose corrections were closest to the mean correction over this period.  This procedure required applying the corrections for a subset of antennas from 8 different datasets, which was performed and loaded into the Telescope Monitor and Control DataBase (TMCDB) on October 26.  The online control software then used these positions for subsequent observations.
A few more antenna moves occurred after this date (DV12, DV16, PM02, PM03, and PM04), requiring additional updates on November 10 and 17 using weighted mean positions derived from several  subsequent all-sky runs (see \S~\ref{tools}). These optimal mean positions were subsequently used by the offline calibration software package -- Common Astronomy Software Applications (CASA) \cite{CASA} -- in both the manual \cite{human} and automated pipelines \cite{hiroko,pipeline}, to retroactively correct the Cycle 3 science data obtained prior to the date when these positions were loaded into the TMCDB.

\section{Results from the weather station network}

\subsection{New ALMA weather stations}

\begin{figure}[h!]    
\centering
\includegraphics[scale=0.80]{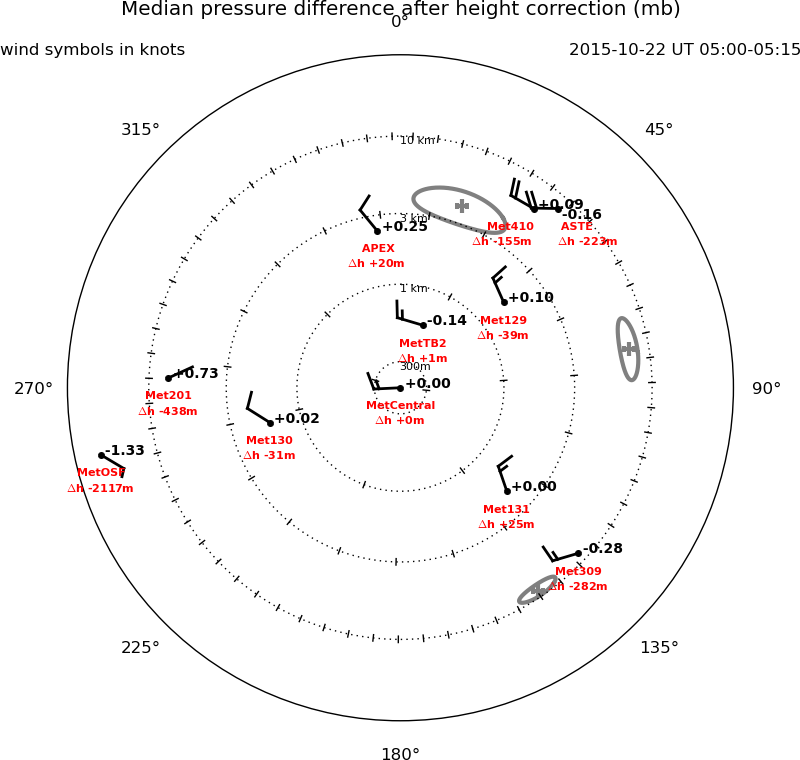}
\caption{A weather map of the Chajnantor area is shown in radially logarithmic polar coordinates during a 15-minute nighttime period on 2015-10-22.  The symbols indicate the wind speed and direction and the adjacent numbers in black give the pressure differential with respect to station MeteoCentral. The numbers in red give the height differential in meters with respect to MeteoCentral.   On the wind symbol radiating from each point, the short flag lines are 5~knots, the long lines are 10~knots, and the azimuth is the wind direction (north = up = 0$^\circ$).  This map can be compared to the daytime map from the same day in Figure~\ref{logradial2}.}
\label{logradial}
\end{figure}

The ALMA weather stations consist of Vaisala PTU300 units containing sensors for temperature, barometric pressure, and relative humidity, and Vaisala WMT52 units for wind direction and velocity.  During August/September 2015, six new weather stations were deployed near existing antenna pads: three at or near the ends of the three branches, and three part way out (see Figure~\ref{logradial}).  During the LBC, there were 7 antennas located near weather stations (see Table~\ref{stationtable}).

\begin{table}[h]
\begin{center}
\caption{Weather stations and their nearest occupied pads and antennas (as of 2015-Nov-11) \label{stationtable}}
\small
\begin{tabular}{ccccccc} 
\hline\hline
Station & Pad & Antenna & Separation & \multicolumn{3}{c}{Geocentric position (nearest 0.1m)}\\
 &              &        &  (m)  & X & Y & Z \\
\hline
Meteo129 & A129 & DV16 & 68 & 2226292.4 & -5440071.2 & -2480490.6\\
Meteo130 & A130 & DV09 & 65 & 2223475.2 & -5440620.3 & -2481822.7\\
Meteo131 & A131 & DA51 & 67 & 2226146.0 & -5439168.0 & -2482751.7\\
Meteo201 & W201 & DA65 & 59 & 2218047.9 & -5442740.5 & -2480988.9\\
Meteo309 & S309 & DV20 & 57 & 2229937.9 & -5435387.8 & -2486806.9\\
Meteo410 & P410 & DA55 & 57 & 2229279.0 & -5440478.3 & -2476637.9\\
MeteoTB2 & A106 & DV13 & 152 & 2225256.8 & -5440258.8 & -2481099.4 \\
MeteoCentral & A101{$^*$} & DV12 & 294 & 2225008.8 & -5440202.7 & -2481447.2\\
\hline
\end{tabular}\\
{$^*$}{The nearest pad is A088 ($\sim$80~m away) but it was not used during the LBC.}
\end{center}
\end{table}

\begin{figure}[h!]   
\centering
\includegraphics[scale=0.39]{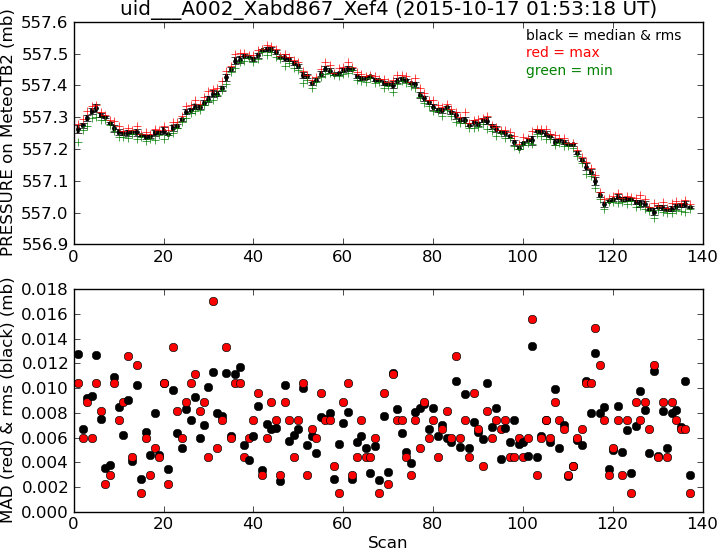}
\includegraphics[scale=0.39]{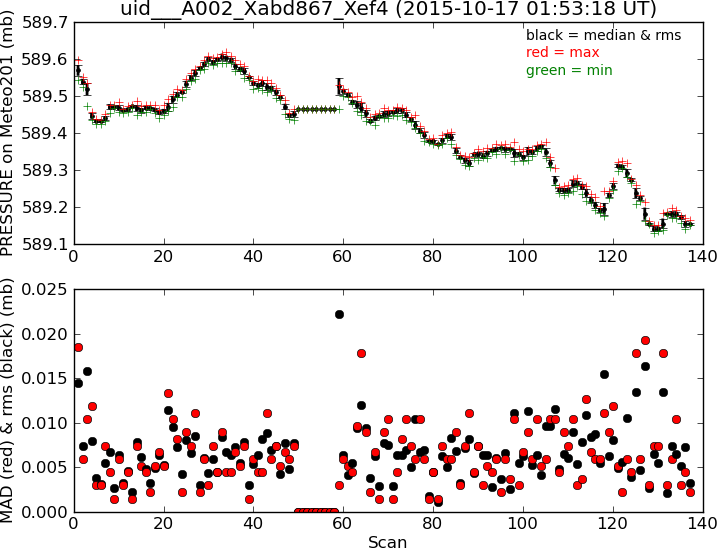}
\includegraphics[scale=0.39]{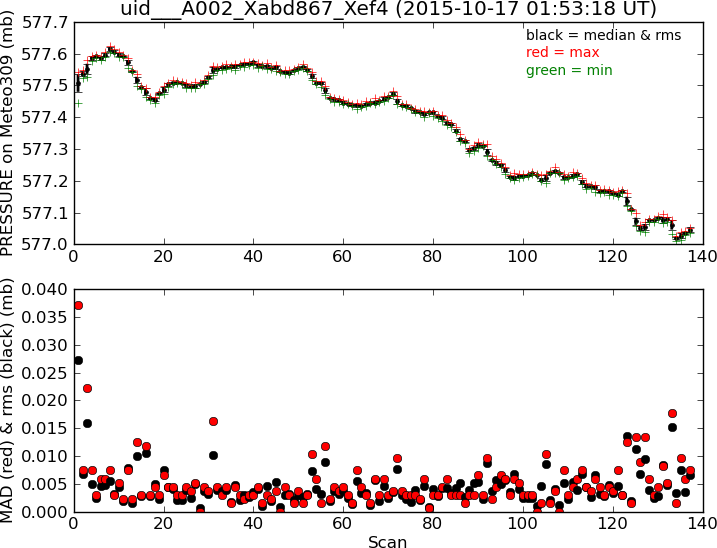}
\includegraphics[scale=0.39]{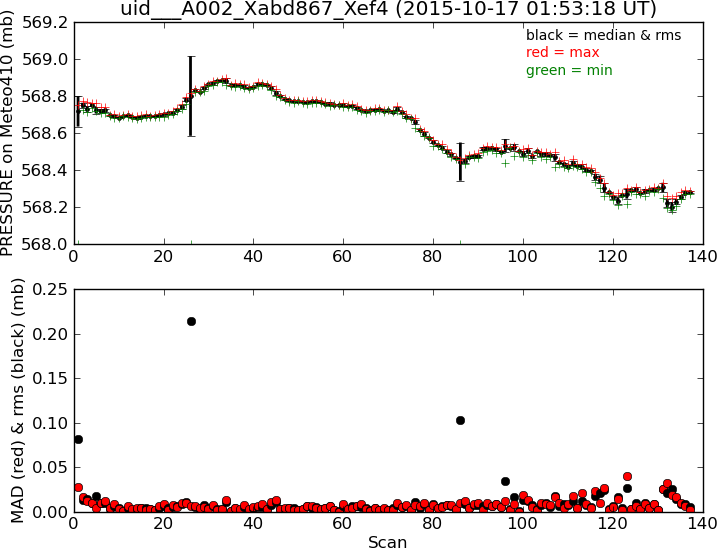}
\caption{Median and median absolute deviation (MAD, scaled to rms) of the weather station barometer 1~Hz values per scan of a Cycle 3 science observation for four of the weather stations (TB2 and the ones at the ends of each branch). Occasional glitches are seen (in this case on station Meteo410), but the MAD is typically 0.005-0.010 mb, which is well below the quoted absolute accuracy.}
\label{weatherstats}
\end{figure}

In Table~\ref{stationtable}, the separation is computed by differencing the pad location with the weather station location stored in the ALMA Science Data Model (ASDM).  The final station became operational at the beginning of 2015 October 16, and they all ran mostly continuously thereafter.
Note that the weather station MeteoTB2 is one of the original weather stations located on the roof of the Array Operations Site (AOS) Technical Building (TB), and served as the sole reference for the delay server throughout ALMA Cycle 3.  We also have the central weather station, MeteoCentral, located near the center of the compact configuration, and the station at the Observatory Support Facility (OSF).  The accuracy of the barometers on these weather stations is remarkable. Many of them were present in the same room at one time at the OSF and all reported the same pressure to within the manufacturer's quoted accuracy, so no special calibration in software was necessary.  The specified repeatability and hysteresis of the Class A pressure sensor are both $\pm0.03$~mb, the linearity is $\pm0.05$~mb, and the calibration uncertainty is $\pm0.07$~mb (all $2\sigma$).  Thus the total root sum square absolute accuracy is $\pm0.10$~mb.  An additional calibration device is being purchased which should further improve the absolute accuracy.  The readback noise on the datastream is also quite low, typically about 0.01 mb rms (Figure~\ref{weatherstats}), although occasional glitches are seen. The manual states that the settling time of the pressure sensor is 2~seconds (100\% response).  All weather quantities are returned in each software poll.

\subsection{Ancillary stations from other facilities}

\begin{figure}[h!]   
\centering
\includegraphics[scale=0.75]{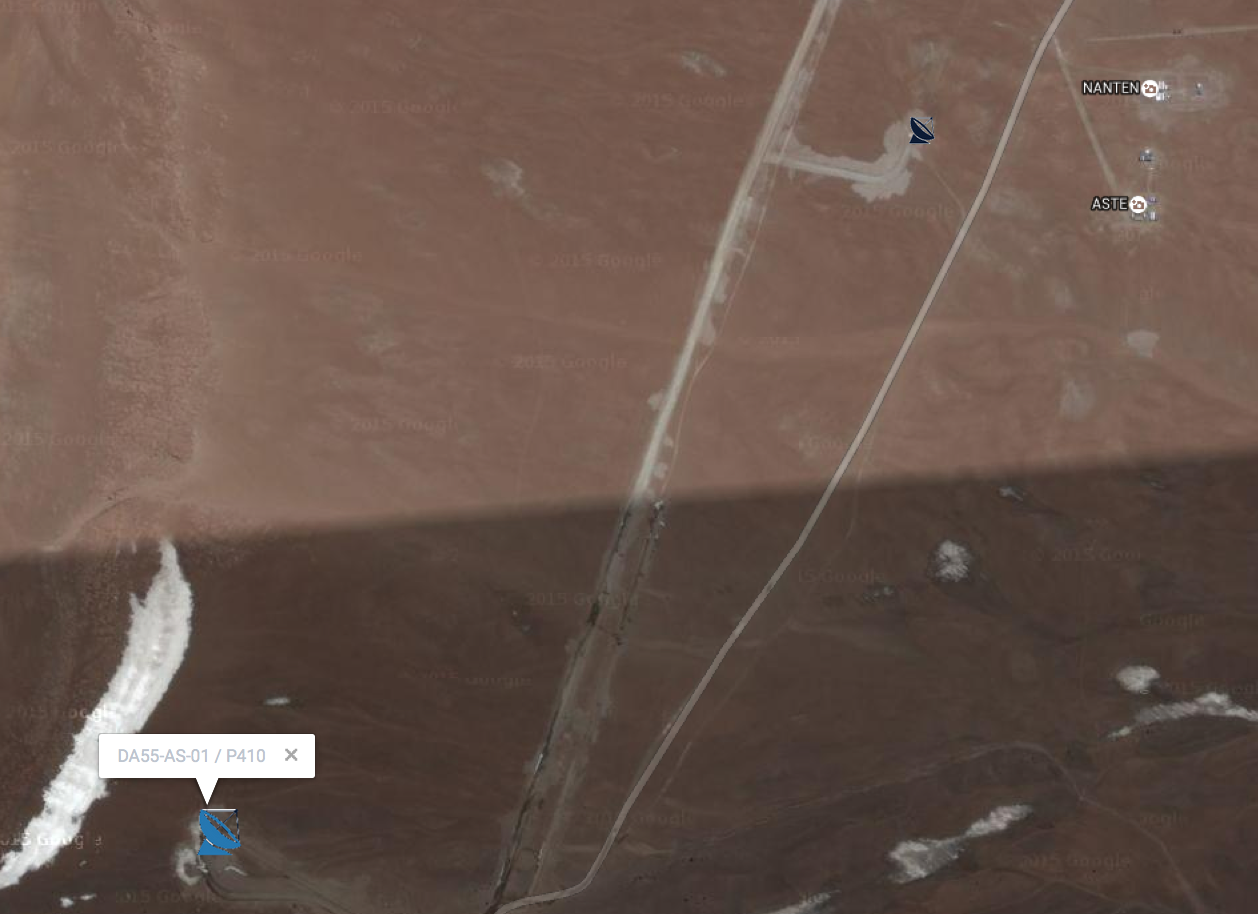}
\caption{Google Earth image showing the proximity of Meteo410 (near antenna pad P410) to the ASTE (1.85~km) and NANTEN2 (1.98~km) facilities.}
\label{aste}
\end{figure}

To augment the ALMA weather data, we retrieved publicly available data from the weather station of the Atacama Pathfinder EXperiment (APEX) telescope \cite{APEX}, and we requested and received data files through November 3 from the Atacama Submillimeter Telescope Experiment (ASTE) station \cite{ASTE}.  The proximity of ASTE and NANTEN2 \cite{Kawamura05} to ALMA pad P410 (and hence Meteo410) is shown in Figure~\ref{aste}.

\begin{figure}[h!]   
\centering
\includegraphics[scale=0.76]{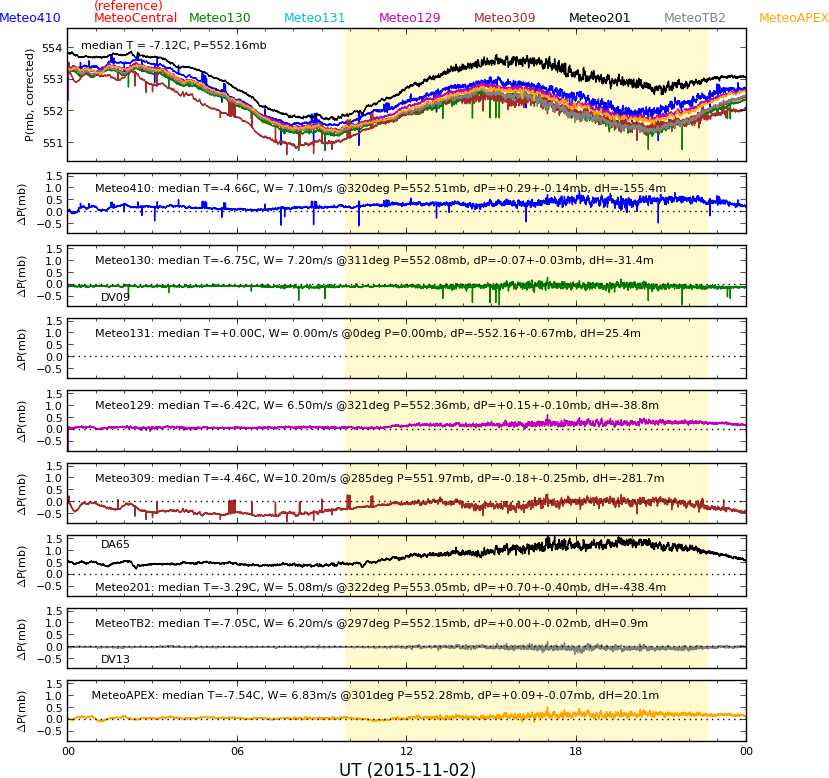}
\caption{Absolute and delta pressure (w.r.t. MeteoCentral) for eight stations vs. time on 2015-11-02 (Meteo131 fiber became non-operational on 2015-10-31). Note the consistent excess pressure seen at Meteo201 and its diurnal variation.}
\label{plotweather}
\end{figure}

\begin{figure}[h!]   
\centering
\includegraphics[scale=0.75]{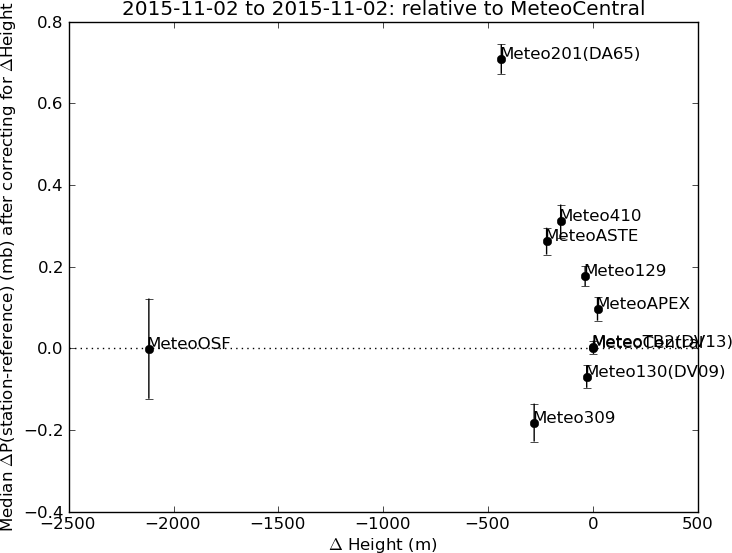}
\caption{The median of the delta pressure (with respect to MeteoCentral) for the 24-hour period of 2015-11-02 after correcting for relative height above the WGS84 ellipsoid for ten stations vs. their relative height above the ellipsoid.  Error bars are the median absolute deviation.  Note the excess positive pressure seen at Meteo201, negative pressure at S309, the good agreement between Pampa la Bola stations P410 and ASTE, and the fact that the OSF pressure agrees rather precisely with the hydrostatic equilibrium formula, in contrast to some of the higher stations.}
\label{dp}
\end{figure}

In order to gather all the data easily, a python script was written ({\tt au.getResampledWeatherFromDatabase}) which can:  (1) retrieve daily weather data at 1~Hz rates from all ALMA stations from the server at weather.aiv.alma.cl, (2) query the APEX station for data, and (3) read from static files containing the ASTE data.  The data are spline interpolated onto a common time grid.  A higher level function ({\tt au.plotWeatherFromDatabase}) calls this function then generates three types of plots: all quantities vs. time, delta pressure vs. time (corrected for height difference using the relative height above the WGS84 ellipsoid \cite{WGS}), delta pressure vs. relative pad height, and a logarithmic radial weather map of delta pressure and wind with the stations in their respective orientation on Chajnantor. Examples are shown for one day in Figures~\ref{logradial}, \ref{plotweather}, and \ref{dp}.  Another higher level function ({\tt au.plotDailyWeather}) generates movies of the logarithmic weather map at a specified cadence (such as 15 minutes).  A movie for the period Oct 16 - Nov 3 can be viewed at \url{http://www.cv.nrao.edu/~thunter/alma/weather.html}.

\section{Exploration of differential barometric pressure}

As soon as we began to collect and analyze the weather data, we noticed that barometric pressure readings from stations near the outer pads at significantly different heights were not consistent with the expected scaling formula with height. This finding challenged the assumption made in the delay server software which uses that formula to compute the pressure at each pad on the basis of the weather station MeteoTB2. Thus we began to investigate the systematic differential pressures measured at the outer stations with respect to MeteoCentral.  

\subsection{Details of the height correction}

\begin{figure}[h!]   
\centering
\includegraphics[scale=0.84]{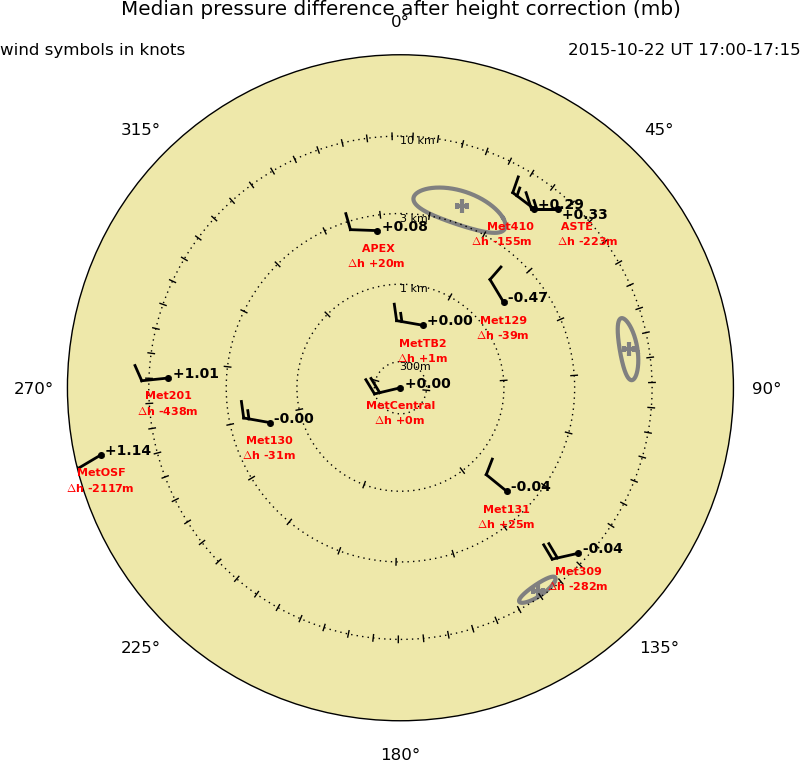}
\caption{Same as Figure~\ref{logradial} but for the daytime (signified by the
yellow background).  Note the winds are higher and more directionally consistent across
 the array and the pressure gradients are more positive than at night.}
\label{logradial2}
\end{figure}

Initially, we were correcting the barometer readings for the relative {\it geocentric} heights of the remote stations compared to the reference station, as does the ALMA delay server when transferring pressure from MeteoTB2 to the remote pads. However, we realized that it should be more accurate to use the height relative to the ellipsoid model of the Earth (WGS84) rather than a sphere.  When we changed the correction to use the ellipsoid, the differential pressures in the P and S branches did indeed decrease significantly in magnitude.  (The W branch pressure differential was essentially unaffected by this change because the sphere and ellipsoid differ only in the north/sound direction.)  Nevertheless, the differentials at the P and S branch stations often remain different from zero. Ideally, we would use the height with respect to the local geoid to perform the height correction, but we are not aware of a tabulation of that surface for our site. In any case, the fact that the sense of the differential -- excess pressure on the west often coupled with a decrement of pressure to the south and east -- is naively consistent with the direction of prevailing winds (from the west or west-northwest), adding support for it being a real effect that we are measuring.  Also, the diurnal dependence of the differential matches the diurnal dependence of the windspeed (i.e. higher daytime winds correspond to higher excess pressure at Meteo 201, see Figure~\ref{logradial2}), lending additional evidence that the effect is real.  

\begin{figure}[h!]   
\centering
\includegraphics[scale=0.165]{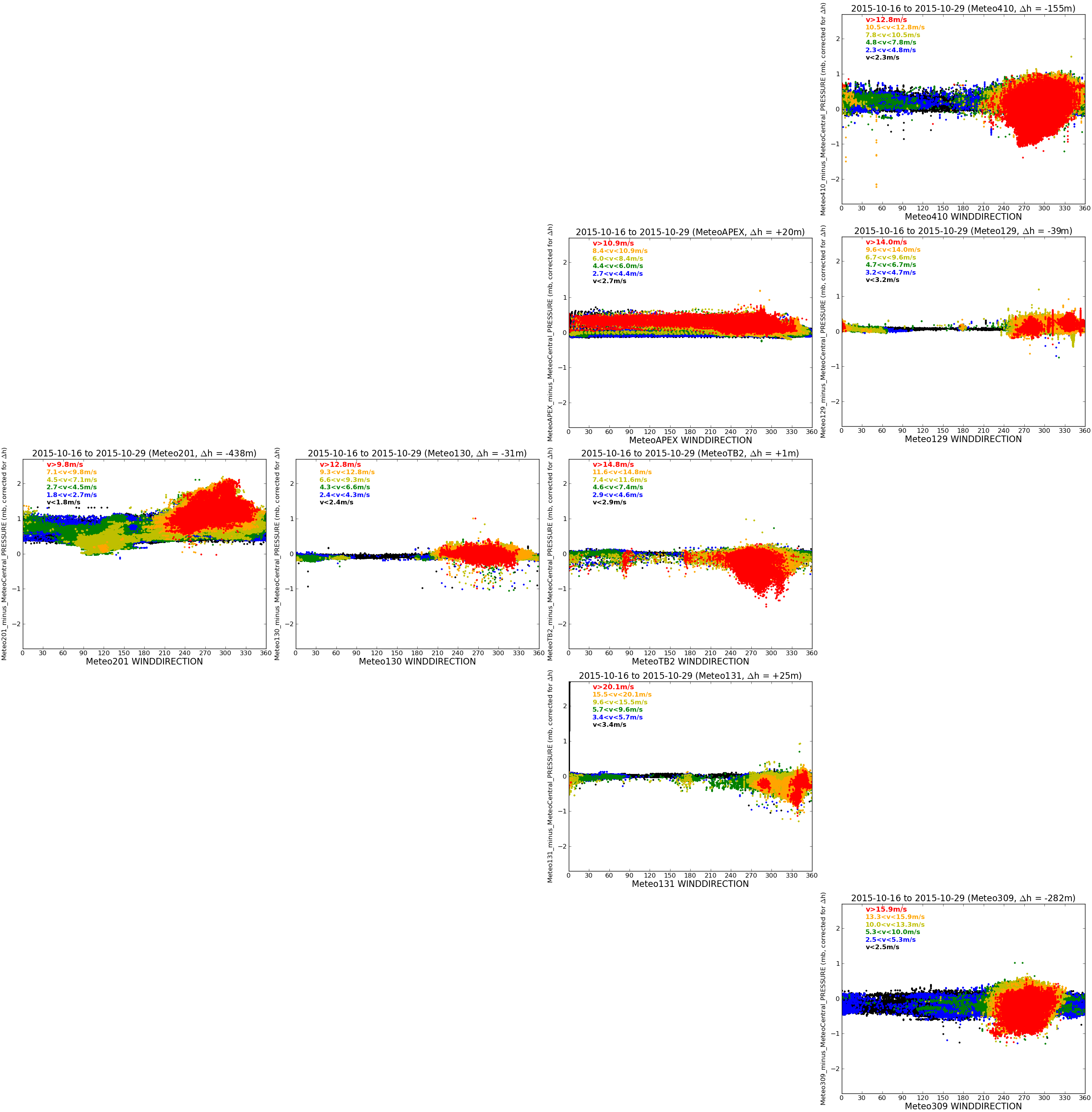}
\caption{Scatter plots of pressure differential vs. wind direction for the 7 ALMA weather stations plus APEX. The x-axis scale is from azimuth 0 to 360 degrees and the y-axis scale is -2.6 to +2.6 mbar. The points are color-coded by wind velocity percentile bins (red: 0-10, orange: 10-25, yellow: 25-50, green: 50-75, blue: 75-90, black: 90-100), with higher velocity points plotted on top of lower velocity points.  The layout of the plots emulates the geographic layout of the stations (north is up, and W201 is to the left).}
\label{colorwinds}
\end{figure}

A summary of the effect can also be visualized in a scatter plot of the excess pressure vs. wind direction where the points are color-coded by wind velocity (Figure~\ref{colorwinds}).  Here we see that when the winds are high, they are invariably coming from the west-northwest according to all of the ALMA weather stations.  The APEX station shows a more broad direction distribution.  At pad W201, when the winds are high, there is a more positive pressure excess; while at pad P410, there is a reduction in the pressure.  Pad S309 is an intermediate case, but slightly favors a negative excess. Interestingly, at W201, we also see a small opposite effect when the wind comes from the east -- the pressure excess turns toward the negative direction.  The presence of this feature further suggests that the origin of the pressure differentials are due to wind interaction with the geography rather than a spurious instrumental effect.

\subsection{Apparent correlation between Z and differential pressure}

\begin{figure}[h!]   
\centering
\includegraphics[scale=0.66]{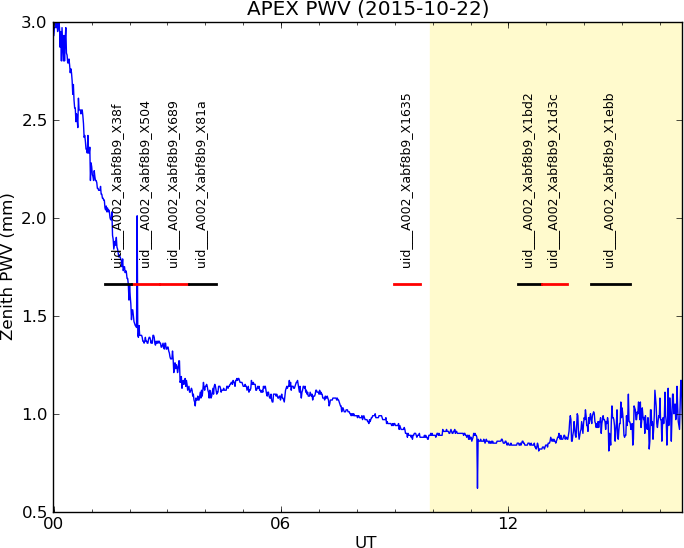}
\includegraphics[scale=0.72]{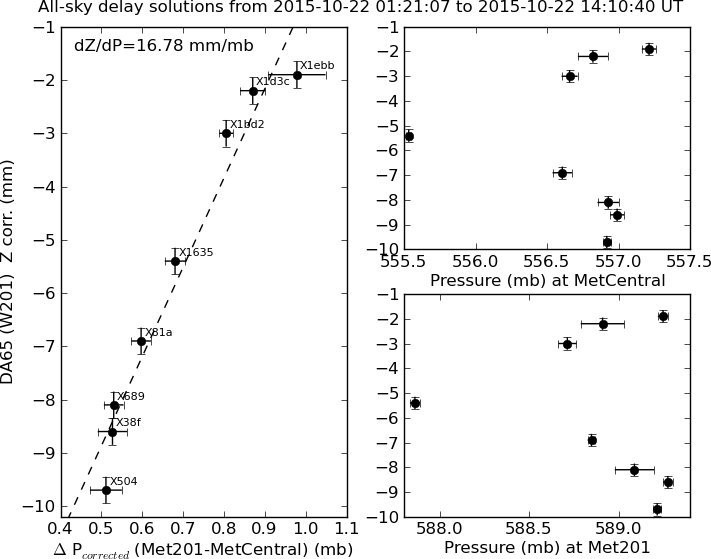}
\caption{Top panel) APEX PWV measured during the morning hours of 2015 October 22.
Lower left panel) The Z correction for DA65 from {\tt tc\_antpos} from a sequence of all-sky quasar observations vs. the differential pressure from Meteo201 compared to MeteoCentral.  The prefix of all the ASDMs is uid://A002/Xabf8b9. The correlation with differential pressure is strong while there is little correlation with the individual station absolute pressures (shown in the two lower right panels).}
\label{zda65}
\end{figure}

On one night (2015 October 22), several all-sky quasar observations were performed during which time the PWV steadily dropped significantly. The median over all antennas dropped from 1.9 to 1.0~mm while the PWV measurement from antenna DA65 (on pad W201, the western-most and lowest antenna location) dropped from 3.0 to 1.4~mm, and the APEX value dropped from 2.0 to 0.9~mm.  The resulting position solution for DA65 changed systematically across nearly 8~mm in the vertical direction.  This change appeared to be well correlated to the differential pressure (Figure~\ref{zda65}), however the slope of a linear fit (16.8~mm/mb) was over 7 times that expected for the excess path due to dry air, which can be estimated from the total excess dry path at the ALMA site (formally computed by performing the integral of dry air density from the surface to the top of the atmosphere in Equation 13.13 of Thompson, Moran \& Swenson \cite{TMS}).  This definite integral simplifies to an expression of the surface pressure ($P_{D,0}$), scale height ($h_{D,0}$), temperature ($T$), molecular weight ($\mathcal{M_D}$), and universal gas constant ($R$):

\begin{equation}
\mathcal{L}_{D} = 0.2228\times10^{-6} \frac{\mathcal{M}_D P_{D,0} h_{D,0}}{RT} = 0.2228\times10^{-6} \frac{(28.964~\rm g/mole)(55600~\rm Pa)(8000~\rm m)}{(8.314~\rm J/mole*K)(273~\rm K)} = 1.26~\rm meters,
\end{equation}

\noindent which is about half the value of at sea level (i.e., $\mathcal{L}_D=2.3$~m at 1013~mb for $h_{D,0}=8000$~m). Thus, the expected effect due to excess dry pressure should be about (1260~mm / 556~mb) = 2.3~mm/mb. We then added a feature to the python script {\tt CompareAntPosResults.py}  which reads the pressure from the nearest weather station and computes the effect of this true pressure measurement on the antenna position and adds a plot of this correction as a fourth row (under the plots of X, Y and Z vs. time and PWV).
Unfortunately, this correction on other nights of all-sky data simply moved the fitted points systematically in the Z direction and did not reduce the scatter.

\subsection{Accounting for the partial pressure of water}

\begin{figure}[h!]   
\centering
\includegraphics[scale=0.75]{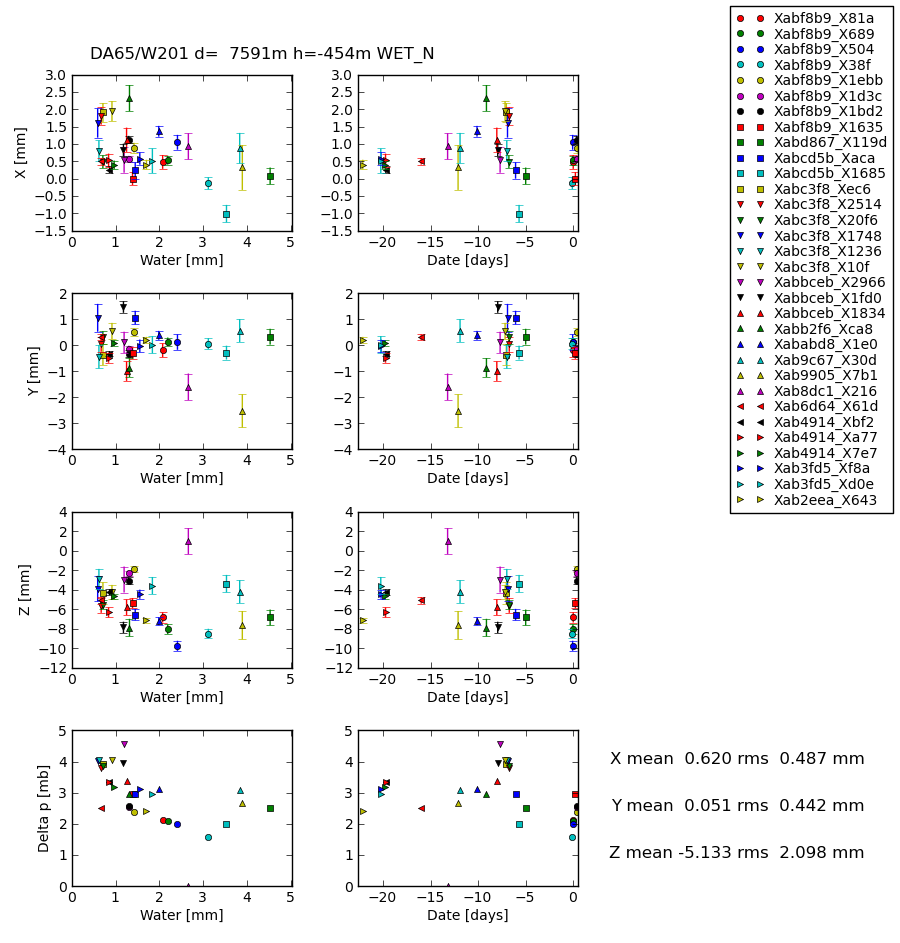}
\caption{Comparison of TelCal-derived antenna position corrections from 32 all-sky observations vs. the PWV for antenna DA65 (top 3 rows) with the pressure correction for that observation (bottom row).  Note the partial corresponding trend between the two lower right panels.  Plot generated by {\tt au.compareAntPosResults}.}
\label{xyzwetNda65}
\end{figure}

We then considered the contribution of the partial pressure of water to the total pressure measured by the barometers.  Since the PWV column varies significantly between the central pads and the distant pads, it is probably more accurate to first remove the partial pressure of water from the reference station pressure measurement to get the dry air pressure, then propagate that pressure to the different height, then restore the partial pressure of water measured at the distant antenna to get the best expected total pressure for comparison to the barometer reading there.  We first considered using the relative humidity from the weather station to determine the water vapor partial pressure, but soon realized that the a better measure of the water content is available from the WVR measurement of the total (path integrated) PWV rather than a value inferred from a local surface reading. We then added this correction to the fourth row of plots produced by {\tt CompareAntPosResults.py}. Initially, some promising agreement was seen in the shape of the correction compared to the shape of the Z axis residual for DA65 (see Figure~\ref{xyzwetNda65}). But ultimately this trend did not repeat on other antennas.

\section{Position updates and the effect of residual errors}

\subsection{Improvements to the software tools for applying position updates}

\label{tools}

\begin{figure}[h!]   
\centering
\includegraphics[scale=0.46]{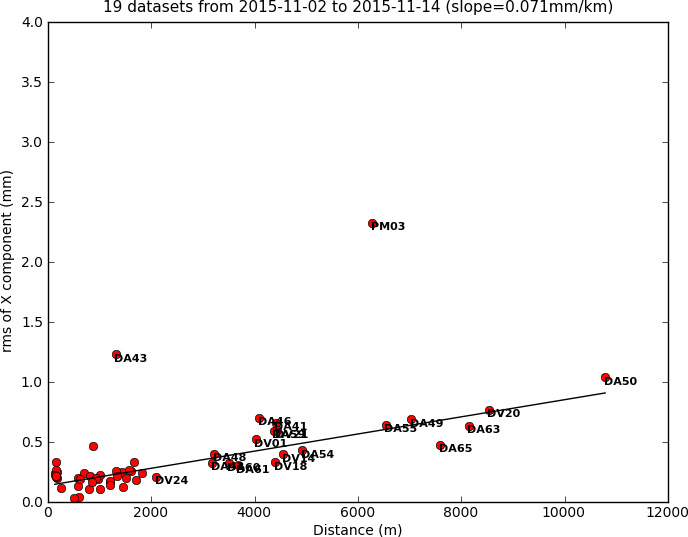}
\includegraphics[scale=0.46]{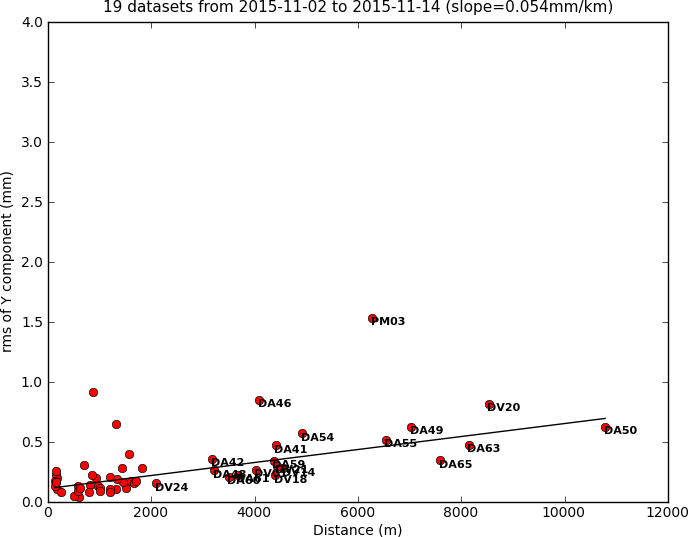}
\includegraphics[scale=0.77]{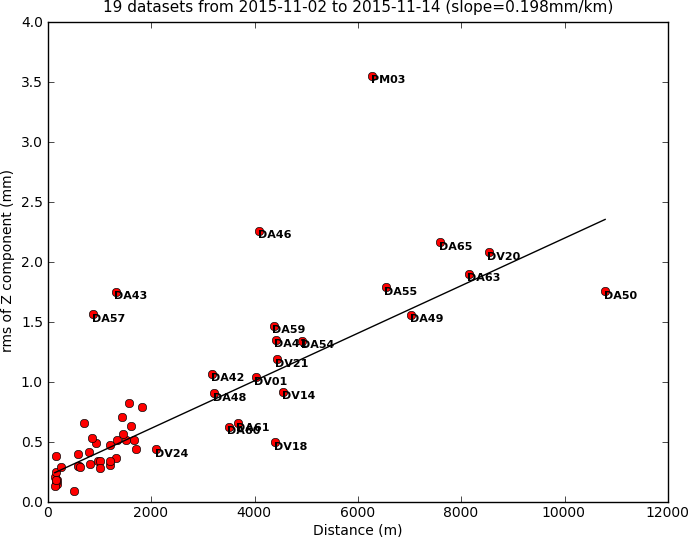}
\caption{The rms of the three components (X=east, Y=north, Z=up) of the antenna position measurements as a function of the distance from the array reference position. The slope in Z is more than 3 times the average of the X and Y slopes.}
\label{rmsvsdistance}
\end{figure}

Traditionally, ALMA antenna position updates have been drawn from the TelCal analysis of a single execution which produces an XML file containing new values. These values are loaded into the TMCDB (using the python script {\tt ImportPositionModel.py}) for any antenna with a total position change that is greater than $50\mu$m and has $>5\sigma$ significance. In order to make it easier to apply the weighted mean of a series of antenna position measurements, we wrote a function {\tt au.averageAntPosResults} to compile a new XML file containing the weighted mean and standard deviation over all the individual antenna position result files gathered in the working directory along with summary plots of each antenna. We also modified the existing tools ({\tt EditPositions.py}, {\tt addAntPos.py} and {\tt UpdatePosition.sh}) to accept additional command line arguments of {\tt -m} (to specify minimum $\sigma$ significance) and {\tt -t} (to specify minimum threshold) to control which antennas to update, and copied the existing option in {\tt EditPositions.py} of {\tt -a} (to select specific antennas to update) to the script ({\tt UpdatePosition.sh}) that updates the revision-controlled file {\tt almaAntPos.txt}, which is ultimately used by CASA calibration scripts to generate {\tt gencal} commands to apply the latest antenna position corrections to the data. These improvements will help keep this file synchronized with the telescope TMCDB which has been difficult to achieve in the past.  Using these new tools, the penultimate position corrections were made on 2015-11-10 for antennas DA61, DV12, DV16, DV19, PM02, PM03, and PM04 using  datasets from Nov 2 through Nov 8 and a threshold of $\sigma > 4$ and $t > 500 \mu$m.  The final corrections of the campaign were made after inspecting the plots of the weighted mean of all datasets recorded from Nov 2 to Nov 14.  This inspection revealed 9 antennas that had a persistent and significant offset from zero.  These antennas were independently identified by choosing $\sigma > 1.6$ and $t >750 \mu$m in {\tt EditPositions.py}, which yielded updates for 9 antennas:  DA52, DA58, DA65, DV01, DV04, DV05, DV12, DV13, DV25, and PM02, which were applied to the TMCDB on 2015-11-17.  The rms of the Z component over those 19 datasets as a function of distance from the array reference position is shown in Figure~\ref{rmsvsdistance}.  There is a steady increase in the dispersion with distance. Dropping four outlier antennas (DA43, PM03, DA46, DA57), the fitted slope in Z (0.2~mm/km) is 3.15 times the mean of the other two coordinates.  Although the determination of the Z coordinate position is inherently more uncertain than the other coordinates (due to the limitations of sky coverage), the fact that the rms in the Z coordinate is more than three times rms in the other coordinates suggests that there may indeed be a remaining systematic error in the delay model. 

\begin{figure}[h!]   
\centering
\includegraphics[scale=0.82]{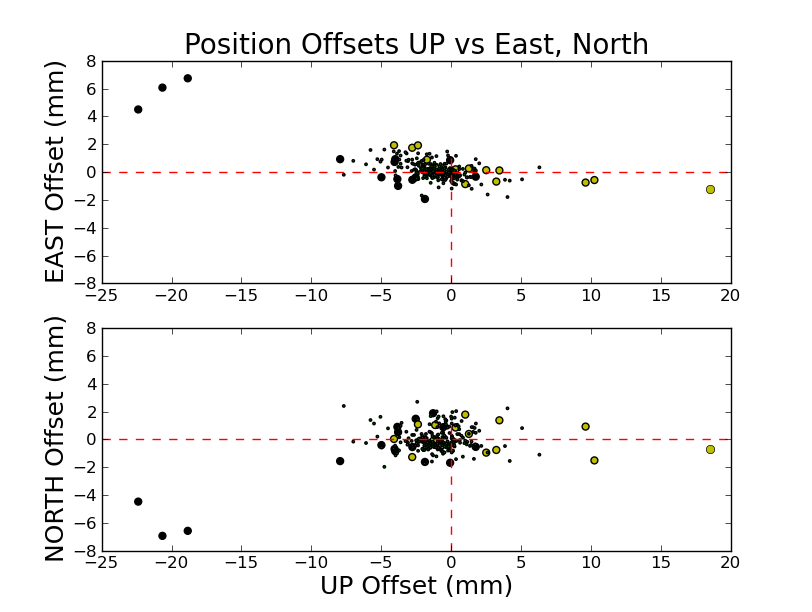}
\caption{Residual antenna position measurements obtained from the 16 allsky observations from 2015/10/14/02:42 to 2015/10/22/14:10.  For each antenna and ENU coordinate axis, the mean has been removed. The scatter in the UP axis is much larger than east or north.  Largest values correspond to the most distant antennas on the western branch (DA65: large black circles) and the northeastern P branch (DA50: large yellow circles).}
\label{eastvsup}
\end{figure}

Another way of comparing the variation in the derived positions between the three axis is shown in Figure~\ref{eastvsup}.  Here, we have removed the average offset for each coordinate of each antenna to show the residual offset.  The plot clearly shows that the overwhelming majority of the baseline changes are in the Up offset.  The slight slope in the East offset is mysterious and may imply a systematic gradient in the troposphere over the array in the east/west direction, perhaps related to the average wind direction.

\subsection{Bottom line for observations}

The linear regression fit of 0.198~mm/km in Figure~\ref{rmsvsdistance} can be compared to the ALMA specification on Phase Center long term stability (technical requirement \#154 \cite{TechReq}, flowing down from science requirement \#280 \cite{SciReq}), which is 217~femtoseconds, or 0.065~mm.  Thus, the repeatability of ALMA interferometric baseline measurements is better than the ALMA specification of 0.075~mm only in the compact configurations ($< 300$~m maximum baseline).  However, it is important to realize that the important quantity is the baseline error times the sine of the separation angle between the science target and the phase calibrator.  Thus, the ALMA operations team have strived to identify nearby calibrators (via the faint calibrator surveys) for each science project in order keep this angle below $3^\circ$ in configurations with maximum baselines of 5~km and larger.  In this case $65\mu$m / sin($3^\circ$) = $1200\mu$m, meaning that a baseline error up to about 1~mm can be acceptable.

\section{Future directions}

The failure to reduce the scatter in the antenna position measurements suggest that the problem is not merely due to not accounting for the differences in dry air pressure at the different antenna pads.  It may be due to our inaccurate modeling of the water vapor above the antennas. There are many reasons to suspect this as our current limitation.

\subsection{WVR sky coupling efficiency and tc\_wvrskydip}

Currently, TelCal assumes a default sky coupling efficiency of the WVR receiver when analyzing WVR data.  Attempts have been made to measure the efficiency on the sky using rapid skydip scans obtained on 2015 August 22, October 22, and October 31.  Initial analysis of these data yielded efficiency values that are larger than 1 in some cases, which is obviously unphysical. If these efficiencies can be accurately measured, then they can be written into the ASDM from the TMCDB so that both online and offline TelCal be able to read them.

\subsection{Oxygen sounder: lapse rate and scale height measurements}

\begin{figure}[h!]   
\centering
\includegraphics[scale=0.94]{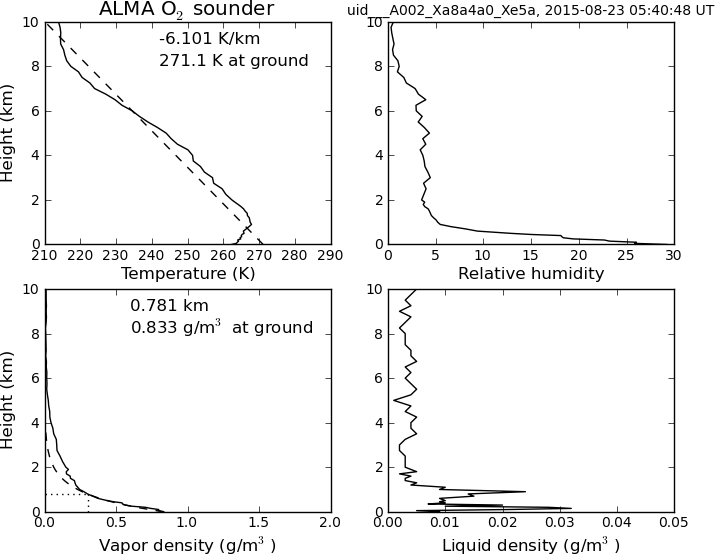}
\caption{The oxygen sounder level 2 data product during one of the WVR skydip calibration datasets. The fits are simple linear regression and locating the 1/e point with respect to the ground vapor density. Plot produced by {\tt au.o2sounder}.}
\label{o2sounderprofiles}
\end{figure}

Currently, TelCal assumes a fixed temperature lapse rate (derivative with respect to height) and water vapor scale height (1~km). For test purposes, ALMA operates a multi-channel 50-60 GHz scanning microwave radiometer purchased from Radiometrics Inc. which retrieves the vertical profile (58 values) for four meteorological quantities at the AOS: water vapor volume density, liquid water volume density, relative humidity and temperature. The data are written to a text file on a linux machine and are not yet included in the individual ASDMs nor the monitor database.
 An example of one set of profiles
produced by the python script {\tt au.o2sounder} is shown in Figure~\ref{o2sounderprofiles}.  A movie of a few days produced by the same script is available at
\url{http://www.cv.nrao.edu/\~thunter/alma/o2sounder.html}. Simple fits to the water vapor scale height and temperature lapse rate for a specific ASDM can be performed using linear regression. Typical values are 0.8-1.2~km and $-6.3$~K/km, however the vapor distribution is not always well characterized by a single exponential and the temperature profile is not really linear, particularly when an inversion layer is present (Figure~\ref{o2sounderprofiles}).   In order to explore the effect of using  the measured lapse rate and scale height in TelCal, we added these as input parameters to the functions in which they were previously hardcoded as constants.  It is now possible to reprocess all the baseline runs obtained on dates when the oxygen sounder data are available, first by running {\tt tc\_wvr}, then {\tt tc\_antpos}.  If this produces more consistent antenna position results, then the next step is to modify TelCal to accept the profiles themselves rather than the linear fits and to get the oxygen sounder profiles automatically written into the ASDM.

\subsection{Populate more of the outer pads, particularly to the west}

\begin{figure}[h!]   
\centering
\includegraphics[scale=0.68]{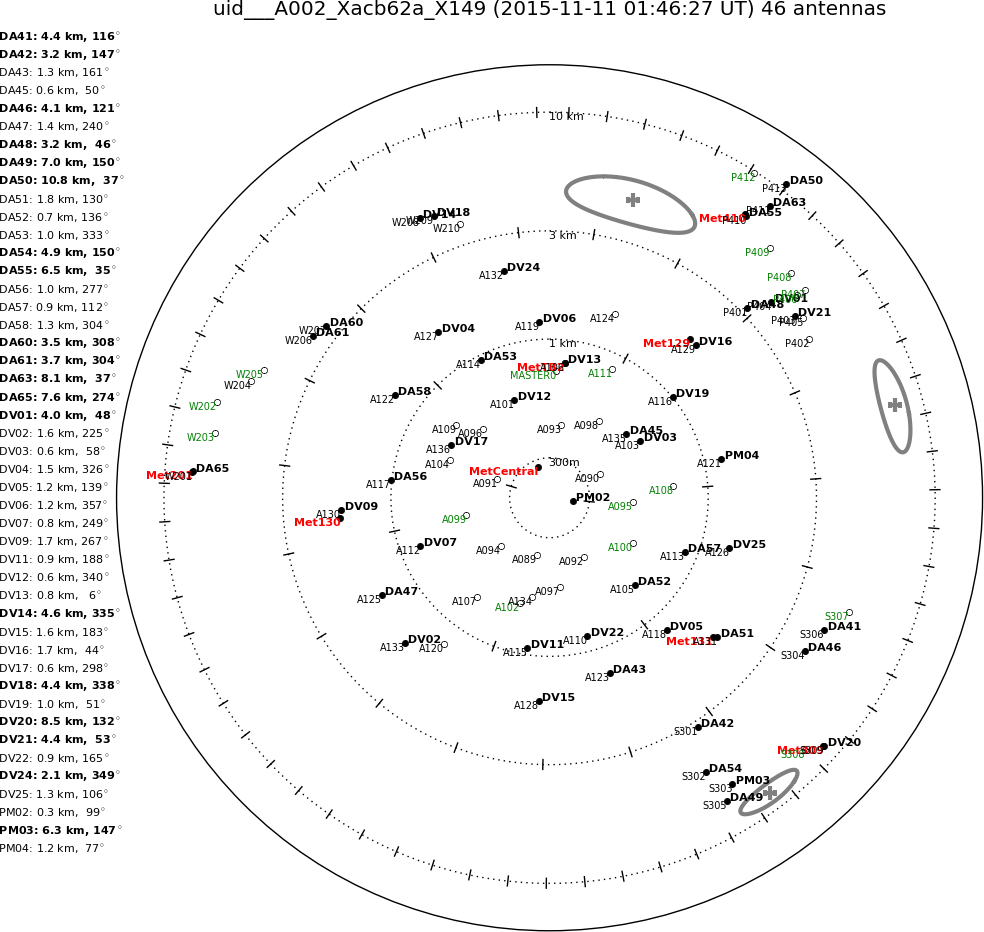}
\caption{A logarithmic radial plot of the 2015 Long Baseline Campaign configuration with the occupied pads shown by filled points, and the unoccupied pads (beyond the inner array) shown by open points. Of the 29 never-occupied pads, the 15 of them outside the inner array are labelled in green font and the weather stations are labeled in red font.  The azimuthal tick marks denote 1~km spacing on each circle.  Grey ellipses mark some of the major mountain peak features.}
\label{allpads}
\end{figure}

Judging by the contents of the history file {\tt antennaMoves.txt}, 29 of the ALMA pads have still never been occupied by an antenna (see Figure~\ref{allpads}).  Pads of particular note are W202, W203 and W205.  In future LBCs (possibly as soon as October 2016), it would be useful to populate some of these pads in addition to W201 as it would provide some shorter baselines (1-2~km) to this pad.  This would better determine how the antenna position measurement uncertainty increases along this branch and may help discern whether the variation mostly due to the height difference or to the distance westward.  

\subsection{Measuring and removing anomalous delay}

If the scatter in antenna position measurements cannot be solved by these efforts, it may still be possible to mitigate their effect by performing much more frequent measurements.  That is, if an accurate set of antenna positions can be obtained once in very dry stable weather, then it may be possible to precede each subsequent science observation (on other nights under other conditions) with a shorter (15~minute) version of the all-sky quasar observations in order to later solve for the residual zenith path delay above each antenna for that observation.  This strategy would only work if this residual delay remains sensibly constant over a typical science observation, which appears to be the case since there is not that much change between successive baseline runs on one night. Of course, the normal temporal phase referencing \cite{Fomalont2014} would still be needed to remove short term variations.



\acknowledgments 

The National Radio Astronomy Observatory is a facility of the National
Science Foundation operated under agreement by the Associated
Universities, Inc. ALMA is a partnership of ESO (representing its
member states), NSF (USA) and NINS (Japan), together with NRC (Canada)
and NSC and ASIAA (Taiwan), and KASI (Republic of Korea), in
cooperation with the Republic of Chile. The Joint ALMA Observatory is
operated by ESO, AUI/NRAO and NAOJ.  We thank Takeshi Okuda for
sending the ASTE weather files covering the period of our analysis.
This research made use of NASA's Astrophysics Data System
Bibliographic Services.

\bibliography{report.bib}{} 

\begin{thebibliography}{10}

\bibitem{Ma98}
C.~{Ma}, E.~F. {Arias}, T.~M. {Eubanks}, A.~L. {Fey}, A.-M. {Gontier}, C.~S.
  {Jacobs}, O.~J. {Sovers}, B.~A. {Archinal}, and P.~{Charlot}, ``{The
  International Celestial Reference Frame as Realized by Very Long Baseline
  Interferometry},'' {\em Astronomical Journal} {\bf 116}, pp.~516--546, July
  1998.

\bibitem{Hills2010}
R.~E. {Hills}, R.~J. {Kurz}, and A.~B. {Peck}, ``{ALMA: status report on
  construction and early results from commissioning},'' in {\em Ground-based
  and Airborne Telescopes III},  {\em Proc. of SPIE} {\bf 7733}, p.~773317,
  July 2010.

\bibitem{Hills08}
R.~E. {Hills} and A.~J. {Beasley}, ``{The Atacama Large
  Millimeter/submillimeter Array},'' in {\em Ground-based and Airborne
  Telescopes II},  {\em Proc. of SPIE} {\bf 7012}, p.~70120N, July 2008.

\bibitem{Marson08}
R.~{Marson}, J.~{Kern}, A.~{Farris}, and R.~{Hiriart}, ``{A dual-consumer
  design for the Atacama Large Millimeter Array control subsystem},'' in {\em
  Advanced Software and Control for Astronomy II},  {\em Proc. of SPIE} {\bf
  7019}, p.~701905, Aug. 2008.

\bibitem{Amestica06}
R.~{Amestica}, B.~{Gustafsson}, and R.~{Marson}, ``{Time synchronization within
  the ALMA software infrastructure},'' in {\em Society of Photo-Optical
  Instrumentation Engineers (SPIE) Conference Series},  {\em Proc. of SPIE}
  {\bf 6274}, p.~627416, June 2006.

\bibitem{Farris05}
A.~{Farris}, R.~{Marson}, and J.~{Kern}, ``{The ALMA Telescope Control
  System},'' in {\em 10th ICALEPCS Int. Conf. on Accelerator \& Large Expt.
  Physics Control Systems. Geneva, 10 - 14 Oct 2005, WE3A.3-6O},  Oct. 2005.

\bibitem{Cherednichenko2010}
S.~{Cherednichenko}, A.~{Emrich}, and T.~{Peacocke}, ``{Water Vapor Radiometer
  for ALMA: Optical Design and Verification},'' in {\em Twenty-First
  International Symposium on Space Terahertz Technology},  p.~389, Mar. 2010.

\bibitem{Emrich09}
A.~{Emrich}, S.~{Andersson}, M.~{Wannerbratt}, P.~{Sobis}, S.~{Cherednichenko},
  D.~{Runesson}, T.~{Ekebrand}, M.~{Krus}, C.~{Tegnader}, and U.~{Krus},
  ``{Water Vapor Radiometer for ALMA},'' in {\em Twentieth International
  Symposium on Space Terahertz Technology},  E.~{Bryerton}, A.~{Kerr}, and
  A.~{Lichtenberger}, eds., pp.~174--177, Apr. 2009.

\bibitem{Nikolic13}
B.~{Nikolic}, R.~C. {Bolton}, S.~F. {Graves}, R.~E. {Hills}, and J.~S.
  {Richer}, ``{Phase correction for ALMA with 183 GHz water vapour
  radiometers},'' {\em A\&A} {\bf 552}, p.~A104, Apr. 2013.

\bibitem{LBC2014}
{ALMA Partnership}, E.~B. {Fomalont}, C.~{Vlahakis}, S.~{Corder}, A.~{Remijan},
  D.~{Barkats}, R.~{Lucas}, T.~R. {Hunter}, C.~L. {Brogan}, Y.~{Asaki}, and
  et~al., ``{The 2014 ALMA Long Baseline Campaign: An Overview},'' {\em ApJL}
  {\bf 808}, p.~L1, July 2015.

\bibitem{Telcal2011}
D.~{Brogui{\`e}re}, R.~{Lucas}, J.~{Pardo}, and J.-C. {Roche}, ``{TELCAL: The
  On-line Calibration Software for ALMA},'' in {\em Astronomical Data Analysis
  Software and Systems XX},  I.~N. {Evans}, A.~{Accomazzi}, D.~J. {Mink}, and
  A.~H. {Rots}, eds., {\em Astronomical Society of the Pacific Conference
  Series} {\bf 442}, p.~277, July 2011.

\bibitem{Telcal2004}
D.~{Brogui{\`e}re}, F.~{Cosson}, H.~{Hafok}, R.~{Lucas}, and J.~{Pardo},
  ``{ALMA On-Line Calibration Software},'' in {\em Astronomical Data Analysis
  Software and Systems (ADASS) XIII},  F.~{Ochsenbein}, M.~G. {Allen}, and
  D.~{Egret}, eds., {\em Astronomical Society of the Pacific Conference Series}
  {\bf 314}, p.~101, July 2004.

\bibitem{SDP81}
{ALMA Partnership}, C.~{Vlahakis}, T.~R. {Hunter}, J.~A. {Hodge}, L.~M.
  {P{\'e}rez}, P.~{Andreani}, C.~L. {Brogan}, P.~{Cox}, S.~{Martin},
  M.~{Zwaan}, and et~al., ``{The 2014 ALMA Long Baseline Campaign: Observations
  of the Strongly Lensed Submillimeter Galaxy HATLAS J090311.6+003906 at z =
  3.042},'' {\em ApJL} {\bf 808}, p.~L4, July 2015.

\bibitem{CASA}
J.~P. {McMullin}, B.~{Waters}, D.~{Schiebel}, W.~{Young}, and K.~{Golap},
  ``{CASA Architecture and Applications},'' in {\em Astronomical Data Analysis
  Software and Systems XVI},  R.~A. {Shaw}, F.~{Hill}, and D.~J. {Bell}, eds.,
  {\em Astronomical Society of the Pacific Conference Series} {\bf 376},
  p.~127, Oct. 2007.

\bibitem{human}
S.~L. {Schnee}, C.~{Brogan}, D.~{Espada}, E.~{Humphreys}, S.~{Komugi},
  D.~{Petry}, B.~{Vila-Vilaro}, and E.~{Villard}, ``{The human pipeline:
  distributed data reduction for ALMA},'' in {\em Observatory Operations:
  Strategies, Processes, and Systems V},  {\em Proc. of SPIE} {\bf 9149},
  p.~91490Z, Aug. 2014.

\bibitem{hiroko}
H.~{Shinnaga}, E.~{Humphreys}, R.~{Indebetouw}, E.~{Villard}, J.~{Kern},
  L.~{Davis}, R.~E. {Miura}, T.~{Nakazato}, K.~{Sugimoto}, G.~{Kosugi},
  E.~{Akiyama}, D.~{Muders}, F.~{Wyrowski}, S.~{Williams}, J.~{Lightfoot},
  B.~{Kent}, E.~{Momjian}, T.~{Hunter}, and {ALMA Pipeline Team}, ``{ALMA
  Pipeline: Current Status},'' in {\em Revolution in Astronomy with ALMA: The
  Third Year},  D.~{Iono}, K.~{Tatematsu}, A.~{Wootten}, and L.~{Testi}, eds.,
  {\em Astronomical Society of the Pacific Conference Series} {\bf 499},
  p.~355, Dec. 2015.

\bibitem{pipeline}
D.~{Muders}, F.~{Wyrowski}, J.~{Lightfoot}, S.~{Williams}, T.~{Nakazato},
  G.~{Kosugi}, L.~{Davis}, and J.~{Kern}, ``{The ALMA Pipeline},'' in {\em
  Astronomical Data Analysis Software and Systems XXIII},  N.~{Manset} and
  P.~{Forshay}, eds., {\em Astronomical Society of the Pacific Conference
  Series} {\bf 485}, p.~383, May 2014.

\bibitem{APEX}
R.~{G{\"u}sten}, L.~{\AA}. {Nyman}, P.~{Schilke}, K.~{Menten}, C.~{Cesarsky},
  and R.~{Booth}, ``{The Atacama Pathfinder EXperiment (APEX) - a new
  submillimeter facility for southern skies -},'' {\em A\&A} {\bf 454},
  pp.~L13--L16, Aug. 2006.

\bibitem{ASTE}
H.~{Ezawa}, R.~{Kawabe}, K.~{Kohno}, and S.~{Yamamoto}, ``{The Atacama
  Submillimeter Telescope Experiment (ASTE)},'' in {\em Ground-based
  Telescopes},  J.~M. {Oschmann}, Jr., ed., {\em Proc. of SPIE} {\bf 5489},
  pp.~763--772, Oct. 2004.

\bibitem{Kawamura05}
A.~{Kawamura}, N.~{Mizuno}, Y.~{Yonekura}, T.~{Onishi}, A.~{Mizuno}, and
  Y.~{Fukui}, ``{NANTEN2: A Submillimeter Telescope for Large Scale Surveys at
  Atacama},'' in {\em IAU Symposium},  {\em IAU Symposium} {\bf 235}, p.~275,
  2005.

\bibitem{WGS}
``Geomatics guidance note 7 part 2: Coordinate conversions and transformations
  including formulas,'' Tech. Rep. 373-7-2, International Association of Oil \&
  Gas Producers, July 2012.
\newblock Section 2.2.1.

\bibitem{TMS}
A.~R. {Thompson}, J.~M. {Moran}, and G.~W. {Swenson}, {\em {Interferometry and
  Synthesis in Radio Astronomy, John Wiley \& Sons, 2007.}}, John Wiley \&
  Sons, 2007.

\bibitem{TechReq}
K.-I. {Morita}, M.~{Sugimoto}, M.~{Miccolis}, D.~{Sramek}, P.~{Napier},
  P.~{Yagoubov}, and N.~{Whyborn}, ``Atacama large millimeter/submillimeter
  array system technical requirements,'' Tech. Rep. ALMA-80.04.00.00-005-C-SPE,
  ALMA, December 2012.

\bibitem{SciReq}
A.~{Wootten} and T.~{Wilson}, ``Atacama large millimeter/submillimeter array
  scientific specifications and requirements,'' Tech. Rep.
  ALMA-90.00.00.00-001-A-SPE, ALMA, July 2006.

\bibitem{Fomalont2014}
E.~{Fomalont}, T.~{van Kempen}, R.~{Kneissl}, N.~{Marcelino}, D.~{Barkats},
  S.~{Corder}, P.~{Cortes}, R.~{Hills}, R.~{Lucas}, A.~{Manning}, and
  A.~{Peck}, ``{The Calibration of ALMA using Radio Sources},'' {\em The
  Messenger} {\bf 155}, pp.~19--22, Mar. 2014.

\end{thebibliography}
\bibliographystyle{spiebib} 

\end{document}